\documentclass[Review,sageh,times]{sagej}

\usepackage[utf8]{inputenc}
\usepackage[T1]{fontenc}
\usepackage{moreverb,url}
\usepackage{xurl}
\usepackage{graphicx}
\usepackage{tabularx,booktabs}
\usepackage{array}
\usepackage{caption}
\usepackage{subcaption}
\usepackage{multirow}
\usepackage[table]{xcolor}
\usepackage{amsmath,amssymb}
\usepackage{microtype}
\usepackage{placeins}
\usepackage{marvosym}
\usepackage[colorlinks,bookmarksopen,bookmarksnumbered,citecolor=blue,urlcolor=blue,linkcolor=blue]{hyperref}

\graphicspath{{figures/}}
\newcolumntype{Y}{>{\centering\arraybackslash}X}
\definecolor{uaIndustrial}{HTML}{D85AD7}
\definecolor{uaDense}{HTML}{D7191C}
\definecolor{uaMedium}{HTML}{FF4A4A}
\definecolor{uaLow}{HTML}{F58B8B}
\definecolor{uaGreen}{HTML}{45B649}
\definecolor{uaPasture}{HTML}{EFE55A}
\definecolor{uaArable}{HTML}{FFF26D}

\begin{document}

\title{Exploring Urban Land Use Patterns by Pattern Mining and Unsupervised Learning}

\author{Zdena Dobesova\affilnum{1} and Tai Dinh\affilnum{2}\textsuperscript{\Letter} and Pavel Novak\affilnum{1}}

\affiliation{\affilnum{1}Department of Geoinformatics, Palacký University, Olomouc, Czech Republic\\
\affilnum{2}The Kyoto College of Graduate Studies for Informatics, 7 Tanaka Monzencho, Sakyo, Kyoto, Japan}

\email{Zdena Dobesova (zdena.dobesova@upol.cz); Tai Dinh (t\_dinh@kcg.ac.jp, it08duytai@gmail.com); Pavel Novak (pavelnovak1996@email.cz)}

\begin{abstract}
Comparative planning needs reproducible methods for identifying recurring land-use configurations across cities. Using Urban Atlas 2018 data for 100 European urban areas, we construct 290,396 focal-neighborhood transactions and 1,543 frequent-itemset support features at 10\% minimum support. Ward clustering is applied in the original normalized feature space, with UMAP used only for visualization. A seven-cluster descriptive solution is retained through multi-criterion evaluation and 500 paired feature-subsampling repetitions. Sensitivity analyses show stable city-similarity geometry across 5--15\% thresholds but greater membership sensitivity to representation choices and non-artificial land-use content. The framework supports peer-city comparison while making scale and boundary limitations explicit.
\end{abstract}


\maketitle

\section{Introduction}\label{sec:introduction}

Urban areas are complex systems influenced by socioeconomic, environmental, and infrastructural factors, with land use patterns being key to urban morphology. Understanding these patterns is crucial for urban planning and policy. The Copernicus program's Urban Atlas \citep{european_environment_agency_data_2025} provides detailed data on land cover, land use, and spatial development across European cities, offering valuable insights into urban landscapes.

Traditional geographic approaches, including spatial analysis, remote sensing, and geographic information systems (GIS), provide powerful tools for examining urban land-use patterns, temporal change, and environmental impacts. However, standardized comparison across large numbers of cities and the identification of recurring multi-category land-use associations can remain analytically challenging. Data-mining approaches can complement these established methods by systematically extracting recurring structures from large and heterogeneous spatial datasets.

In recent years, geography has increasingly adopted data mining and machine learning to analyze large datasets and uncover patterns. These data-driven approaches provide complementary tools for identifying recurring structures and associations in large and complex datasets. Among the diverse array of data-mining techniques, frequent itemset mining (FIM) and clustering are particularly relevant to the systematic comparison of recurring land-use combinations across cities.

Existing comparative studies have characterized cities using street-network structure, spatial metrics, land-use mix, urban morphology, and aggregate city typologies \citep{louf_typology_2014,ye_quantitative_2014,prastacos_using_2017,gregor_similarities_2018}. These approaches provide valuable descriptions of urban composition and form, but they do not directly characterize how multi-category land-use combinations recur across local neighborhoods and how the prevalence of these combinations differs systematically between cities. This leaves a planning-relevant gap in standardized cross-city comparison: recurring local land-use co-presence can provide information that complements aggregate measures of composition, density, and land-use mix, but its comparative structure and sensitivity to analytical choices remain insufficiently examined. Such neighborhood-level comparisons can support the identification of comparable urban contexts, unusual land-use interfaces, and candidate peer cities for more detailed planning investigation.

This study addresses this gap by representing each city through the relative support of recurring local combinations of Urban Atlas land-use classes. In this study, a \emph{pattern} refers to an unordered set of land-use categories that are co-present within the specified local neighborhood, rather than to their exact geometric arrangement. \emph{Similarity} refers to proximity between cities based on their profiles of relative support for these recurring patterns. The resulting clusters are therefore interpreted as descriptive empirical groupings conditional on the data representation and analytical choices, rather than as fixed or normative urban typologies.

Accordingly, the study addresses the following research questions:

\begin{enumerate}
\item What recurring local land-use co-presence configurations characterize the selected European urban areas?
\item How do cities differ in the prevalence of these configurations, and which urban-form and urban--rural interface characteristics distinguish the resulting descriptive city groupings?
\item How can these recurring configurations support comparative planning and urban benchmarking, and how sensitive are the resulting comparisons to the support threshold, analytical representation, clustering choices, and non-artificial land-use content?
\end{enumerate}

The study makes four contributions. First, it develops a reproducible procedure for transforming polygon-based Urban Atlas data into transaction data representing recurring local co-presence of land-use classes. Second, it uses frequent itemset mining to construct comparable city-level profiles of the prevalence of these recurring land-use combinations across 100 selected European urban areas. Third, it compares cities by clustering their normalized frequent-itemset support profiles directly in the original feature space and evaluates the resulting structure through cluster-number assessment and extensive sensitivity analyses. Fourth, it examines how support thresholds, alternative representations and clustering geometries, UMAP visualization settings, repeated transaction patterns, and non-artificial land-use content affect the comparison, thereby defining both the robustness and the limitations of the resulting descriptive groupings for comparative planning.

The analytical framework comprises four main stages: (1) preprocessing Urban Atlas spatial data and converting local land-use neighborhoods into transactions; (2) extracting frequent itemsets using the negFIN algorithm \citep{aryabarzan_negfin_2018}; (3) constructing a city-by-itemset matrix in which each city is represented by the relative support of its frequent itemsets; and (4) comparing cities using hierarchical agglomerative clustering and sensitivity analysis. Ward hierarchical clustering is performed on the normalized original city-by-itemset feature matrix. UMAP \citep{mcinnes_umap_2018} is used only for two-dimensional visualization and diagnostic assessment and does not determine the primary cluster assignments. Heatmaps, dendrograms, and UMAP projections are subsequently used to interpret and visualize the resulting city groupings.

The datasets, source code, analysis notebook, and computational workflow used in this study are publicly available through the ULUMiner GitHub repository (\url{https://github.com/taiduydinh/ULUMiner}) to support reproducibility and reuse. The repository contains the materials required to reproduce the results reported in this study, together with information on data provenance and the analytical workflow.

The remainder of this paper is organized as follows. The Related Work section reviews the relevant literature and establishes the planning and methodological context of the study. The Preliminaries, Data and Proposed Method section describes the data and analytical framework. The Experimental Results section presents the empirical findings and sensitivity analyses. The Discussion interprets the findings in relation to comparative planning and considers the methodological limitations of the approach. The Conclusion summarizes the main findings and directions for future research.

% -----------------------------------------------------------------------------

\section{Related Work}\label{sec:related-work}
\subsection{Comparative urban form, land-use mix, and city typologies}

The comparative study of urban structure has a long history. Classical work described recurring configurations of urban areas and their evolution \citep{harris_nature_1945,harris_nature_1997}, while more recent studies have used increasingly detailed spatial data to quantify differences in urban form. For example, commuting patterns have been used to examine monocentric and polycentric urban structures \citep{burger_heterogeneous_2011}, while street-network characteristics have supported cross-city comparison based on block form, orientation, configuration, and entropy \citep{louf_typology_2014,boeing_measuring_2018,boeing_urban_2019,boeing_multi-scale_2020}. These approaches demonstrate that cities can be compared systematically through measurable characteristics of their spatial structure.

Among recent approaches to the quantitative characterization of urban form, \citet{fleischmann2026hierarchical} introduced the Hierarchical Morphotope Classification (HiMoC), a theory-driven and computationally scalable framework for classifying built form. HiMoC operationalizes the concept of the morphotope---the smallest locality with a distinctive morphological character---by delineating contiguous, morphologically distinct areas and organizing them into a hierarchical taxonomic structure based on morphometric characteristics derived primarily from building footprints and street networks. The framework demonstrates how detailed morphological classification can be extended consistently across large geographic extents while retaining an explicit connection to urban-morphological theory. Its focus, however, is on the morphology of the built environment rather than on the recurrence and prevalence of multi-category land-use co-presence, which constitutes the complementary perspective examined in the present study.

Land-use composition and mixture provide another important perspective on urban form. \citet{ye_quantitative_2014} combined space syntax, density measures, and a mixed-use index to examine relationships among street-network integration, building density, and land-use mixture, while \citet{hoek_towards_2010} used mixed-use measures to compare urban areas in the Netherlands. Using Urban Atlas data, \citet{prastacos_using_2017} applied spatial metrics such as polygon size, edge length, polygon density, shape, and diversity to characterize urban areas. At a broader European scale, the European Environment Agency developed a typological approach based on land-use and socioeconomic characteristics to compare a large set of European cities and support differentiated urban-policy perspectives \citep{gregor_similarities_2018}. Together, these studies show the value of standardized indicators and typologies for comparative urban analysis, but they primarily characterize cities through aggregate composition, morphology, network structure, or summary indicators.

\subsection{Data-driven representations and clustering of urban structure}

Machine-learning and data-mining approaches have expanded the range of representations available for comparing urban form. Previous work using Urban Atlas data represented city-center extracts as images and used features derived from pretrained neural networks to identify similar European cities through clustering \citep{dobesova_similarity_2019,dobesova_experiment_2020,dinh2025efficient}. More recent approaches represent urban morphology through multiple spatial scales and data structures. For example, \citet{chen_developing_2023} introduced a multi-scale representation integrating building-, block-, and neighborhood-level characteristics, while \citet{gui_novel_2025} developed an adaptive multi-scale framework for capturing variation in urban morphology across heterogeneous urban environments. Related research increasingly combines raster, topological, graph-based, and multi-scale representations to characterize urban form \citep{ren_urban_2024,gui_novel_2025,chen_interpreting_2025}.

Clustering remains a common tool for identifying recurring urban structures once such representations have been constructed. For example, \citet{bautista-hernandez_jobs-housing_2024} used hierarchical clustering to distinguish employment- and housing-oriented zones in Mexico City, while \citet{wu_machine_2024} clustered street-based local areas across multiple cities to identify urban--suburban and polycentric structural patterns. \citet{zhang_image-based_2025} similarly used hierarchical clustering of urban road-network representations and related the resulting groups to geographic and socioeconomic characteristics. These studies illustrate the usefulness of unsupervised learning for comparative urban analysis, while also showing that the interpretation of clusters depends strongly on how urban structure is represented before clustering.

\subsection{Comparative planning, institutional context, and urban--rural interfaces}
The physical form of cities is shaped not only by land-use composition and morphological processes but also by planning institutions, regulatory frameworks, and governance arrangements. In a cross-national analysis, \citet{schmidt_planning_institutions_2024} show that national planning frameworks differ in characteristics such as restrictiveness, legal bindingness, and the balance between national and local orientation, and that these institutional differences are associated with patterns of metropolitan spatial development. This finding is particularly relevant to cross-city comparison because similar spatial configurations need not arise from identical planning processes, while cities operating under different institutional systems may nevertheless exhibit comparable physical forms. Consequently, empirical city groupings should not be interpreted directly as planning-system types without independent institutional evidence.

At the same time, urban morphology provides an important empirical basis for planning analysis. 
\citet{gu_teaching_urban_design_2020} emphasizes the role of morphological knowledge in understanding and managing urban landscape change and in connecting analysis of physical form with planning and urban design practice. Similarly, \citet{oliveira_urban_morphology_2022} presents urban morphology as the study of streets, blocks, plots, buildings, and the processes and actors through which these elements are transformed. At a broader comparative scale, \citet{schwarz_urban_form_2010} analyzed urban-form indicators for 231 European cities and demonstrated substantial variation among cities, emphasizing that such differences should be considered when comparing urban areas or developing planning policies across Europe. These studies support the use of standardized spatial evidence for comparative planning while also emphasizing the need to interpret quantitative similarities within their local and institutional contexts.

Comparative indicators and benchmarking can also support planning by identifying meaningful similarities, differences, and candidate cases for more detailed investigation. For example, \citet{monteiro_benchmarking_cities_2024} demonstrate how quantitative urban-form indicators can be used to benchmark alternative city configurations and provide information relevant to planning decisions. Data-driven comparison has also entered planning research more directly. \citet{brinkley_nlp_plans_2024}, for example, used natural language processing to compare 461 city general plans, illustrating how computational methods can facilitate large-scale comparison while retaining a policy-relevant interpretation. In the present study, benchmarking has a more limited and descriptive meaning: cities are compared according to recurring land-use co-presence profiles, rather than ranked according to planning performance or assumed to share the same planning institutions.

The urban--rural interface is another important consideration for interpreting city form. \citet{allen_periurban_interface_2003} argues that rural and urban characteristics frequently coexist within and beyond urban boundaries and that a simple urban--rural dichotomy is inadequate for planning in peri-urban contexts. More recently, \citet{hibbard_rurality_planning_2024} highlight the persistent marginalization of rurality within planning scholarship and emphasize the importance of context-sensitive approaches to rural and regional planning. These perspectives are especially relevant to comparative land-use analysis because official urban boundaries may contain substantial non-artificial land and because urban--rural interfaces can constitute meaningful components of urban structure rather than merely unwanted noise in the data. Accordingly, sensitivity to non-artificial land-use content and boundary definition is necessary when translating empirical city similarities into planning interpretations.

\subsection{Spatial frequent-itemset mining and the research gap}

Frequent itemset mining (FIM), originally associated with market-basket analysis, identifies sets of items that repeatedly occur together within transactions \citep{shekhar_frequent_2017}. Spatial applications extend this principle by treating spatially related phenomena as co-occurring items. Early work on spatial association rules demonstrated how frequent patterns could be derived from relationships among geographic features such as water bodies, roads, and mines \citep{koperski_discovery_1995}. FIM has subsequently been applied to other geographic problems, including recurring road-accident configurations \citep{geurts_understanding_2005} and relationships among spatial variables associated with urban sprawl \citep{pampoore-thampi_mining_2021}. These applications demonstrate that frequent patterns can reveal recurring combinations that are not captured by considering variables independently.

However, the literature reviewed above leaves a specific gap for comparative urban analysis. Existing cross-city studies predominantly characterize urban areas using aggregate land-use composition, spatial metrics, street-network properties, image-derived features, or other morphology indicators. Spatial frequent-itemset studies, in contrast, demonstrate the ability to identify recurring co-occurrences but have not been widely used to construct standardized city-level profiles of recurring multi-category land-use co-presence for comparison across a large set of cities. Consequently, limited evidence exists on how cities differ in the prevalence of such local combinations, how these differences relate to broader urban-form and urban--rural interface characteristics, and how sensitive the resulting city groupings are to analytical choices such as minimum support, representation, clustering geometry, and non-artificial land-use content.

This study addresses that gap by using frequent itemset support profiles to represent recurring local land-use co-presence across 100 selected European urban areas and by comparing those profiles in the original feature space. The resulting groupings are treated as descriptive tools for comparative analysis and urban benchmarking rather than as fixed or normative city typologies. This framing complements established composition-, morphology-, and network-based approaches by focusing specifically on the recurrence of local multi-category land-use combinations and by explicitly examining the sensitivity and limitations of the resulting cross-city comparison.

\section{Preliminaries, Data and Proposed Method}\label{sec:method}

\begin{figure}[!htb]
\centering
\includegraphics[width=\linewidth]{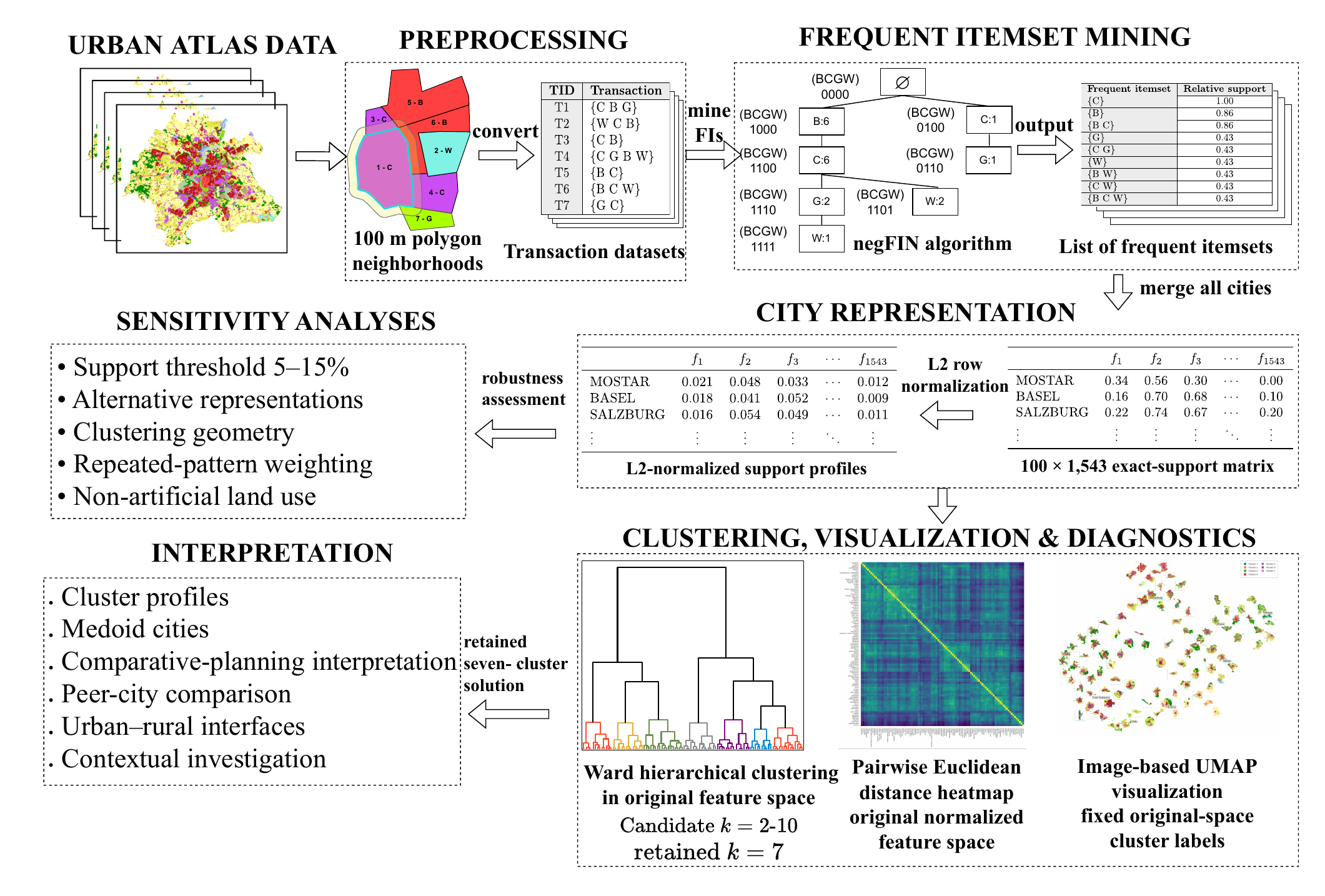}
\caption{Workflow of the proposed comparative land-use framework. Urban Atlas polygons from 100 selected European urban areas are converted into focal-neighborhood transactions and mined for frequent itemsets. Cities are represented by L2-normalized exact-support profiles and clustered using Ward hierarchical clustering directly in the original 1,543-dimensional feature space. Candidate cluster numbers are evaluated before retaining a seven-cluster descriptive solution. Sensitivity analyses assess robustness to alternative analytical choices, while the pairwise-distance heatmap and UMAP provide diagnostic and visual representations. UMAP is used only for visualization and displays the fixed original-space cluster labels. Cluster profiles and medoid cities support comparative-planning interpretation and identification of candidate peer cases.}
\label{fig:workflow}
\end{figure}

Figure~\ref{fig:workflow} illustrates the main analytical workflow together with the associated sensitivity, visualization, diagnostic, and interpretation components. Detailed descriptions of the analytical steps are provided in the subsequent subsections.

\subsection{Urban Atlas as the Source Data}\label{sec:urban-atlas}

The Urban Atlas (UA) is a high-resolution land-cover and land-use product of the Copernicus Land Monitoring Service that provides harmonized spatial information for Functional Urban Areas (FUAs) across Europe and the United Kingdom. FUAs comprise urban centers together with their functionally connected commuting zones and provide a standardized spatial basis for comparative urban analysis. Urban Atlas data support applications including land-use monitoring, urban-form comparison, and assessment of spatial change \citep{european_environment_agency_data_2025}.

\begin{figure}[!htb]
\centering
\begin{minipage}[b]{0.43\linewidth}
\centering
\includegraphics[width=\linewidth]{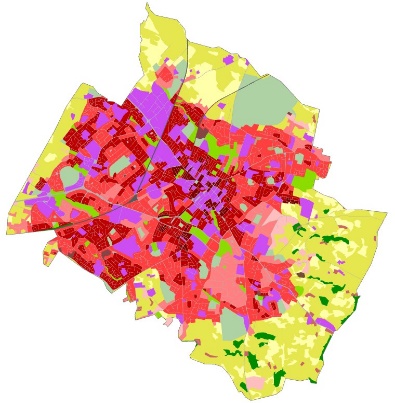}
\end{minipage}\hfill
\begin{minipage}[b]{0.49\linewidth}
\centering
\includegraphics[width=\linewidth]{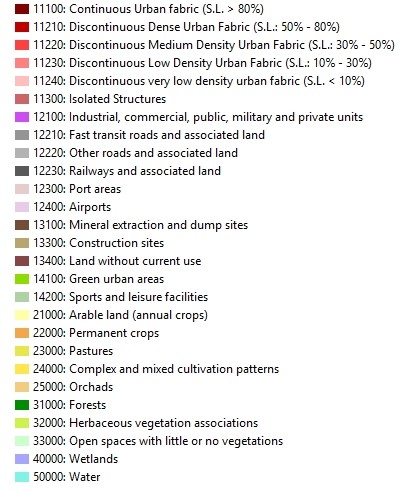}
\end{minipage}
\caption{Example of Urban Atlas land-cover and land-use data for Cheltenham (UK), together with the land-use codes and classes used in the study.}
\label{fig:urban-atlas-cheltenham}
\end{figure}

This study uses the Urban Atlas land-cover and land-use dataset for the 2018 reference year. The dataset is provided as vector polygons and applies a minimum mapping unit of 0.25~ha for urban classes and 1~ha for rural classes \citep{european_environment_agency_mapping_2020}. The 2018 nomenclature distinguishes 27 land-cover and land-use classes within five broad thematic groups: artificial surfaces, agricultural areas, natural and semi-natural areas, wetlands, and water. The classes are represented by hierarchical five-digit codes, which form the categorical basis for the transaction representation used in this study. Figure~\ref{fig:urban-atlas-cheltenham} illustrates the Urban Atlas classification for Cheltenham (UK).

For each selected city, the analysis uses the Urban Atlas Urban Core rather than the complete FUA commuting-zone extent. The Urban Core defines the spatial study extent from which local land-use neighborhoods are constructed before frequent-itemset mining. Because Urban Core boundaries may include both artificial and non-artificial land-use classes, the influence of non-artificial land-use content on cross-city comparison is evaluated in the sensitivity analysis and considered when interpreting the resulting city groupings.

The empirical analysis comprises 100 selected European urban areas. Cities were selected using a population range of 50,000--300,000 inhabitants, with additional consideration given to geographic coverage and representation across European countries. The resulting set is a purposive comparative sample rather than a statistically representative sample of all European FUAs. Consequently, the city groupings are interpreted as descriptive of the analyzed sample, and broader generalization to European cities should be made cautiously.

\subsection{Conversion of Spatial Data into Transaction Data}\label{sec:transactions}

\begin{figure}[!htb]
\centering
\begin{minipage}[t]{0.34\linewidth}
\centering
\includegraphics[width=\linewidth,height=2.05in,keepaspectratio]{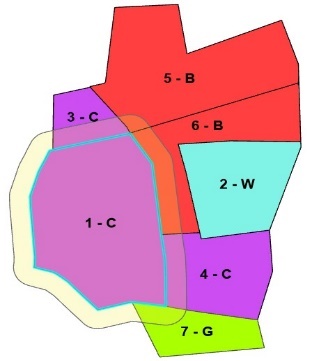}
\captionof{figure}{The zone for finding neighbors for polygon C.}
\label{fig:buffer-zone}
\end{minipage}\hfill
\begin{minipage}[t]{0.60\linewidth}
\centering
\includegraphics[width=\linewidth,height=2.15in,keepaspectratio]{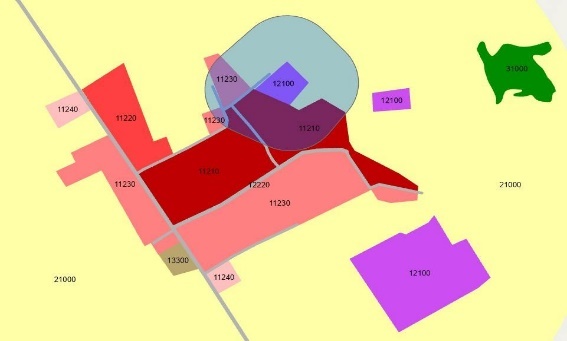}
\captionof{figure}{Testing data of a small village.}
\label{fig:test-village}
\end{minipage}
\end{figure}

The Urban Atlas polygon data were converted into transaction data by treating each polygon as a focal spatial unit and recording the land-use classes co-present within a local buffer around it. A buffer distance of 100~m was used for the primary 100-city analysis. Figure~\ref{fig:buffer-zone} illustrates the construction for a small example in Stetovice, and the corresponding transactions are shown in Table~\ref{tab:neighbors}. One transaction is generated for each focal polygon; therefore, for a fixed study extent, the number of transactions is determined by the number of focal polygons, whereas the buffer distance affects the composition and length of those transactions.

For the selected polygon (No. 1 in Figure~\ref{fig:buffer-zone}), the focal land-use class (C, commercial) is first recorded. Land-use classes of polygons intersecting the buffer are then identified and combined with the focal class. Because a transaction is represented as a set, each land-use code can occur at most once within a transaction: if several neighboring polygons have the same class, that class is recorded once. The final transaction is therefore an unordered set of distinct land-use classes; for example, \{W C B\} and \{B C W\} are equivalent. Table~\ref{tab:neighbors} illustrates this operation for seven focal polygons, producing seven transactions. Removing repeated codes within an individual transaction does not remove repeated transaction rows across different focal polygons; recurring neighborhood configurations therefore continue to contribute repeatedly to itemset support.

\begin{table}[!htb]
\caption{The neighbors for each polygon.}
\label{tab:neighbors}
\centering
\small
\begin{tabularx}{\linewidth}{@{}c c Y Y Y@{}}
\toprule
Polygon no. & Input type of polygon & All neighbors & Neighbors without duplicity & Final transaction \\
\midrule
1 & C & \{C B B C G\} & \{C B G\} & \{C B G\} \\
2 & W & \{C B B\} & \{C B\} & \{W C B\} \\
3 & C & \{C B B\} & \{C B\} & \{C B\} \\
4 & C & \{C G B W\} & \{C G B W\} & \{C G B W\} \\
5 & B & \{B C C\} & \{B C\} & \{B C\} \\
6 & B & \{B C C C W\} & \{B C W\} & \{B C W\} \\
7 & G & \{C C\} & \{C\} & \{G C\} \\
\bottomrule
\end{tabularx}
\end{table}

This transaction representation preserves which distinct land-use classes are co-present within each local neighborhood and how frequently particular combinations recur across focal polygons. It does not preserve the number of neighboring polygons belonging to the same class, polygon area or class-specific coverage within the buffer, polygon geometry, shared-boundary or contact length, contact type or orientation, exact within-buffer distance or overlap extent, or spatial order and configuration. Frequent itemsets derived from these transactions should therefore be interpreted as recurring unordered land-use co-presence within the specified neighborhood, rather than as measures of within-neighborhood dominance, geometric intensity, or exact spatial arrangement.

ArcGIS Pro provides tools such as Polygon Neighbors, Neighborhood Summary Statistics, and Generate Near Table for identifying spatial relationships among polygons, but these tools do not directly produce the transaction format required for FIM. We therefore developed the \emph{SearchDistinctLanduse\_SpatialJoin} model using ArcGIS ModelBuilder and Python. Urban Atlas land-use types are represented by their numeric codes (e.g., 11210 and 11230). The model is illustrated using the Stetovice example near Olomouc in Figure~\ref{fig:test-village} \citep{novak_comparison_2023}.

\begin{figure}[!htb]
\centering
\includegraphics[width=\linewidth]{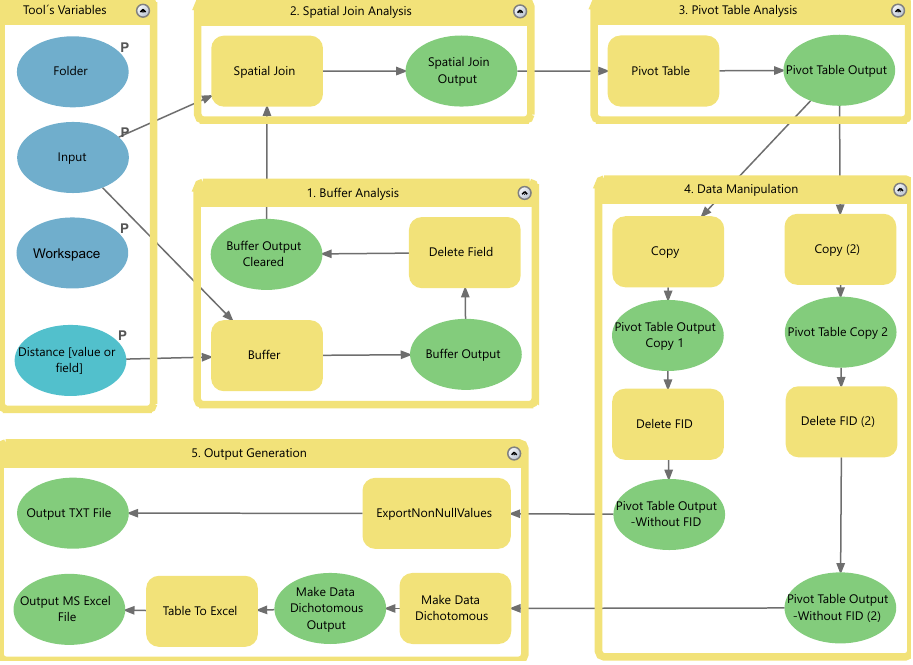}
\caption{Model \emph{SearchDistinctLanduse\_SpatialJoin} with workflow to generate transactions from spatial data.}
\label{fig:model-workflow}
\end{figure}

The \emph{SearchDistinctLanduse\_SpatialJoin} workflow (Figure~\ref{fig:model-workflow}) consists of defining the input parameters, creating the buffer around each focal polygon, performing a spatial join with the Urban Atlas polygon layer, transposing the joined classes through a pivot operation, and generating the transaction outputs \citep{novak_comparison_2023}. A custom Python step writes the transactions to a text file without source-polygon identifiers, while a second output stores the transactions in dichotomous Excel format. The ModelBuilder workflow was also configured for batch processing so that the same procedure could be applied consistently across the selected cities.

City size is not an explicit input to the transaction-construction model. Instead, cities can contain different numbers of focal polygons and hence different numbers of transactions. Cross-city comparison is subsequently based on relative itemset support, defined in the next subsection, so that an itemset is represented by its prevalence among a city\textquotesingle s transactions rather than by its raw support count alone.

Across the 100 study areas, the transaction-construction procedure yielded 290,396 focal-polygon transactions representing 86,591 distinct within-city transaction patterns. Repeated neighborhood configurations were retained: 203,805 transaction occurrences repeat a pattern occurring elsewhere within the same city and therefore continue to contribute to frequent-itemset support. The transactions contain 27 observed Urban Atlas land-use codes. These quantities distinguish recurrence across focal neighborhoods, which is preserved, from repeated occurrences of the same class inside an individual neighborhood, which are collapsed by the set-based representation.

\subsection{Frequent Itemset Mining by negFIN Algorithm}\label{sec:negfin}

Frequent itemset mining (FIM) identifies sets of items that recur within a transaction database. The task was introduced in the association-rule literature by \citet{agrawal_fast_1994} and has since been applied across many domains \citep{luna_frequent_2019}. In the present study, the items are Urban Atlas land-use codes and the transactions are the focal-polygon neighborhoods defined in the preceding conversion subsection.

Given a set of items $I=\{i_1,i_2,\ldots,i_m\}$ and a transaction database $D=\{T_1,T_2,\ldots,T_n\}$, each transaction $T_q$ is a subset of distinct items from $I$. For an itemset $X\subseteq I$, absolute support is the number of transactions containing $X$:
\begin{equation}
\operatorname{sup}(X)=\left|\{T\mid X\subseteq T\land T\in D\}\right|.
\end{equation}
Relative support is $\operatorname{rel\_sup}(X)=\operatorname{sup}(X)/n$. An itemset is frequent when its support meets the selected minimum-support threshold. Relative rather than absolute support is used for cross-city comparison because cities contain different numbers of focal polygons and therefore different numbers of transactions.

\begin{table}[!htb]
\caption{The list of FIs detected from the transaction dataset with \emph{minsup} = 3.}
\label{tab:frequent-itemsets}
\centering
\begin{tabular}{@{}c c c c@{}}
\toprule
FIs & Support & FIs & Support \\
\midrule
\{C\} & 7 & \{W\} & 3 \\
\{B\} & 6 & \{B W\} & 3 \\
\{B C\} & 6 & \{C W\} & 3 \\
\{G\} & 3 & \{B C W\} & 3 \\
\{C G\} & 3 & & \\
\bottomrule
\end{tabular}
\end{table}

For the illustrative transactions in Table~\ref{tab:neighbors}, a minimum absolute support of three yields the itemsets listed in Table~\ref{tab:frequent-itemsets}. For the 100-city analysis, frequent itemsets were mined with negFIN \citep{aryabarzan_negfin_2018}, which represents frequent-itemset search using NegNodesets and a set-enumeration strategy. The primary analysis uses an exact relative minimum support of 10\% within each city. Across the 100 cities, this threshold produces 17,847 nonzero city--itemset support values and a union of 1,543 distinct itemsets. An itemset absent from a city is assigned support zero. All 1,543 resulting features, including land-use code 11100, are retained in the primary city representation.

\subsection{City Representation, Original-Space Clustering, and UMAP Visualization}\label{sec:umap-hac}

The city-by-itemset matrix contains 100 rows and 1,543 columns. Each entry is the exact relative support of one itemset in one city. Table~\ref{tab:city-fis} shows an illustrative extract of this representation.

\begin{table}[!htb]
\caption{Illustrative extract of the city-by-itemset matrix. Each column represents one frequent itemset, and each cell gives its relative support in the corresponding city.}
\label{tab:city-fis}
\centering
\scriptsize
\resizebox{\linewidth}{!}{%
\begin{tabular}{@{}lrrrrcrrc@{}}
\toprule
 & \{11100\} & \{11210\} & \{11220\} & \{11230\} & $\cdots$ & \{11100,12100,12220\} & $\cdots$ \\
\midrule
SHKODER   & 0.34 & 0.56 & 0.30 & 0.48 & $\cdots$ & 0    & $\cdots$ \\
SALZBURG  & 0.22 & 0.74 & 0.67 & 0.33 & $\cdots$ & 0.20 & $\cdots$ \\
INNSBRUCK & 0.22 & 0.59 & 0.62 & 0.32 & $\cdots$ & 0.19 & $\cdots$ \\
MOSTAR    & 0.27 & 0.79 & 0.48 & 0.35 & $\cdots$ & 0.24 & $\cdots$ \\
KORTRIJK  & 0.26 & 0.48 & 0.55 & 0.47 & $\cdots$ & 0.24 & $\cdots$ \\
OOSTENDE  & 0.63 & 0.58 & 0.34 & 0.13 & $\cdots$ & 0.46 & $\cdots$ \\
PLOVDIV   & 0.69 & 0.50 & 0.12 & 0    & $\cdots$ & 0.50 & $\cdots$ \\
BASEL     & 0.16 & 0.70 & 0.68 & 0.36 & $\cdots$ & 0.10 & $\cdots$ \\
BERN      & 0    & 0.46 & 0.67 & 0.60 & $\cdots$ & 0    & $\cdots$ \\
LEMESOS   & 0.58 & 0.70 & 0.53 & 0.36 & $\cdots$ & 0.40 & $\cdots$ \\
$\vdots$  & $\vdots$ & $\vdots$ & $\vdots$ & $\vdots$ & $\ddots$ & $\vdots$ & $\ddots$ \\
PRISTINA  & 0.48 & 0.72 & 0.58 & 0.42 & $\cdots$ & 0.39 & $\cdots$ \\
\bottomrule
\end{tabular}}
\end{table}

Before clustering, each city row is normalized to unit L2 norm so that comparison emphasizes the shape of the itemset-support profile rather than its overall magnitude. Ward hierarchical agglomerative clustering \citep{ward_hierarchical_1963} is then applied directly to this normalized 1,543-dimensional feature matrix using Euclidean geometry. No dimensionality-reduced coordinates or precomputed square distance matrix are used to determine the primary cluster assignments. Pairwise Euclidean distances in this normalized original feature space are used only for diagnostic visualization and comparison.

Candidate cluster numbers $k=2,\ldots,10$ are assessed using four complementary criteria: Silhouette coefficient, Calinski--Harabasz index, Davies--Bouldin index, and feature-subsampling stability \citep{dinh2019estimating,dinh2025data}. Stability is evaluated with 500 paired repetitions. In each repetition, 80\% of the itemset features are sampled without replacement, the sampled rows are L2-normalized, one Ward hierarchy is fitted, and every candidate $k$ is evaluated on that same sampled feature subset. Stability is measured by the adjusted Rand index (ARI) between the full-feature partition and the corresponding feature-subsampled partition. The four criteria are converted to ranks in their appropriate directions and combined by an equal-weight mean rank. This procedure is used to select a defensible descriptive resolution rather than to imply that one value of $k$ is uniquely correct.

UMAP \citep{mcinnes_umap_2018} is used only after the original-space clustering has been defined. Its role is two-dimensional visualization and diagnostic assessment, not cluster construction. The primary display uses $n_{\mathrm{neighbors}}=15$, $\mathrm{min\_dist}=0.1$, Euclidean distance, and random seed 42. Each city is represented by a thumbnail of its Urban Atlas land-use map, with the thumbnail border indicating the fixed original-space Ward cluster. To improve legibility where thumbnails would otherwise overlap, small display-only adjustments are applied to the thumbnail positions while preserving their association with the corresponding UMAP locations. These adjustments affect only the visualization and have no role in the clustering or quantitative UMAP diagnostics.

\subsection{Sensitivity and Diagnostic Analyses}\label{sec:sensitivity-methods}

Sensitivity is evaluated along four dimensions. First, frequent itemsets are re-mined exactly at minimum-support thresholds of 5\%, 7.5\%, 10\%, 12.5\%, and 15\%. At a fixed $k=7$, each alternative threshold is compared with the 10\% solution using ARI, normalized mutual information (NMI), and Spearman correlation between all pairwise city distances.

Second, representation and clustering assumptions are varied by testing integer-rounded support, binary itemset presence, absolute support counts with L2 normalization, exact support without row normalization, cosine distance with average linkage, PCA representations retaining 90\% and 95\% variance followed by Ward clustering, exclusion of itemsets containing road code 12220, restriction to artificial-surface itemsets only, and a one-vote-per-unique-transaction-pattern representation. These are sensitivity variants of the primary analysis: the primary 1,543-feature representation retains road code 12220 as well as both artificial and non-artificial land-use classes. In the artificial-only sensitivity test, artificial surfaces are defined as the observed Urban Atlas codes beginning with 1; consequently, itemsets containing agricultural classes (codes beginning with 2), natural and semi-natural classes (3), wetlands (4), or water (5) are excluded. These tests distinguish sensitivity to support magnitude, repeated neighborhood occurrences, dimensional representation, linkage geometry, roads, and non-artificial land-use content.

Third, UMAP robustness is evaluated over 54 diagnostic embeddings formed by $n_{\mathrm{neighbors}}\in\{5,15,30\}$, $\mathrm{min\_dist}\in\{0,0.1,0.5\}$, Euclidean or cosine distance, and random seeds 0, 42, and 2026. Each embedding is assessed by trustworthiness, overlap of the nearest 10 other cities with the original feature space, Spearman correlation of pairwise distances with the original space, and ARI/NMI between a two-dimensional Ward partition and the fixed original-space partition.

Fourth, the influence of non-artificial land use is quantified at the transaction level using (i) mean number of non-artificial classes per transaction, (ii) share of transactions containing at least one non-artificial class, and (iii) share containing only non-artificial classes. Cluster differences in these measures are summarized descriptively with Kruskal--Wallis tests and Holm-adjusted $p$-values. Because these measures and the cluster labels derive from the same underlying transactions, these tests are treated as exploratory rather than independent confirmatory evidence. Country composition is likewise examined descriptively using Cram\'er's $V$ with a 5,000-permutation test; no causal inference about national planning systems is drawn from this association.

\section{Experimental Results}\label{sec:results}

\subsection{Buffer Scale and Transaction Breadth}\label{sec:buffer}

Buffer distance controls the spatial extent represented by each focal-polygon transaction. It does not change the number of focal polygons for a fixed study extent; instead, a larger buffer can intersect more neighboring polygons and therefore broaden the set of land-use classes represented in each transaction. The scale calibration considered different buffer distances for Olomouc and Enschede, with detailed transaction-length statistics available across 10--200~m for Olomouc.

\begin{figure}[!htb]
\centering
\includegraphics[width=0.80\linewidth,trim=0 0 0 25,clip]{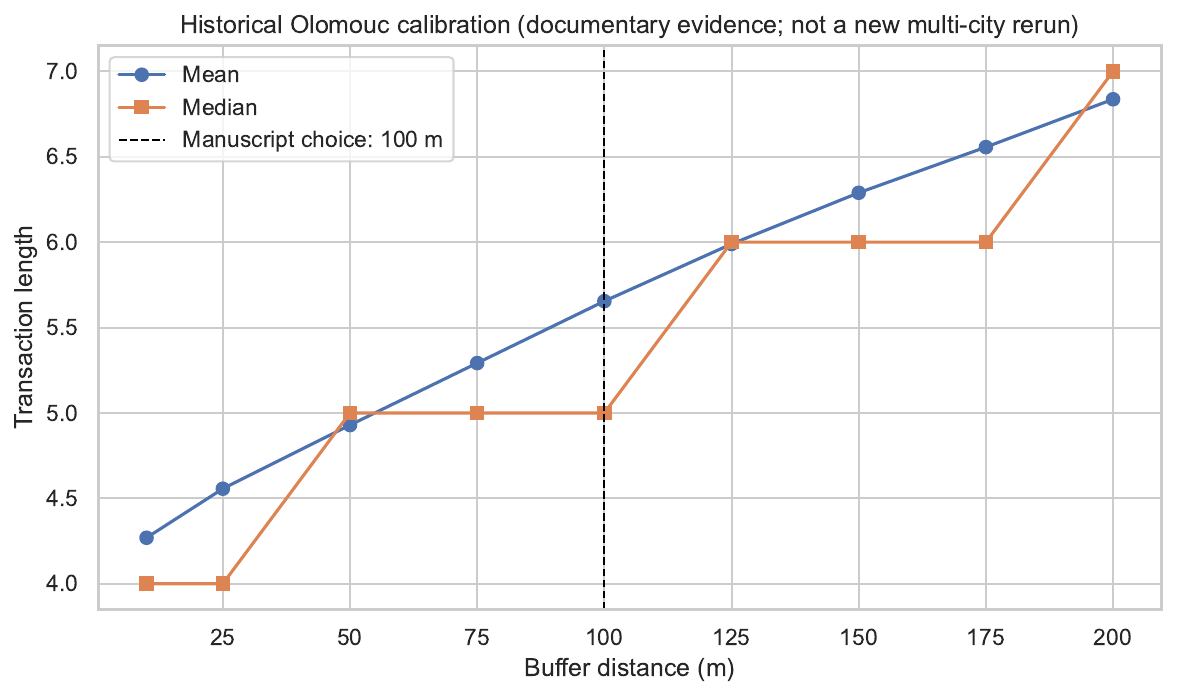}
\caption{Effect of buffer distance on mean and median transaction length for Olomouc. Increasing the buffer broadens the categorical composition represented in focal-polygon transactions. The dashed line indicates the 100~m buffer used in the primary 100-city analysis.}
\label{fig:buffer-calibration}
\end{figure}

In Olomouc, mean transaction length increased from 4.27 land-use classes at 10~m to 6.84 at 200~m, while the median increased from 4 to 7. At the 100~m distance used in the 100-city analysis, the mean and median transaction lengths were 5.65 and 5, respectively (Figure~\ref{fig:buffer-calibration}). The 100~m setting is therefore treated as the selected local analytical scale for this study, not as a universally optimal buffer for heterogeneous European cities. The absence of equivalent multi-city transaction data at alternative buffer distances limits stronger claims about cross-city buffer robustness.

\subsection{Interpreting Frequent Itemsets: Cheltenham}\label{sec:cheltenham}

Cheltenham provides a local example of how frequent itemsets summarize recurring land-use co-presence. At the 10\% minimum-support threshold, exact mining yields 127 frequent itemsets for Cheltenham. Figure~\ref{fig:cheltenham-fis} reports selected multi-category itemsets with comparatively high support; single-category itemsets and several road-dominated combinations are omitted from the display for interpretive clarity, but they remain part of the full mined representation used in the city-level analysis.

\begin{figure}[!htb]
\centering
\small
\begin{tabular}{ccccc}
\toprule
\textbf{Relative support} & \textbf{Absolute support} & \textbf{Land use 1} & \textbf{Land use 2} & \textbf{Land use 3} \\
\midrule
60 & 714 & \cellcolor{uaIndustrial}12100 & \cellcolor{uaDense}\color{white}11210 & \\
56 & 663 & \cellcolor{uaDense}\color{white}11210 & \cellcolor{uaMedium}11220 & \\
54 & 644 & \cellcolor{uaIndustrial}12100 & \cellcolor{uaMedium}11220 & \\
45 & 539 & \cellcolor{uaIndustrial}12100 & \cellcolor{uaDense}\color{white}11210 & \cellcolor{uaMedium}11220 \\
28 & 331 & \cellcolor{uaMedium}11220 & \cellcolor{uaLow}11230 & \\
28 & 332 & \cellcolor{uaDense}\color{white}11210 & \cellcolor{uaGreen}14100 & \\
27 & 322 & \cellcolor{uaMedium}11220 & \cellcolor{uaGreen}14100 & \\
24 & 289 & \cellcolor{uaPasture}23000 & \cellcolor{uaArable}21000 & \\
24 & 288 & \cellcolor{uaIndustrial}12100 & \cellcolor{uaLow}11230 & \\
23 & 275 & \cellcolor{uaDense}\color{white}11210 & \cellcolor{uaMedium}11220 & \cellcolor{uaGreen}14100 \\
23 & 277 & \cellcolor{uaIndustrial}12100 & \cellcolor{uaDense}\color{white}11210 & \cellcolor{uaGreen}14100 \\
\bottomrule
\end{tabular}
\caption{Selected FIs for Cheltenham town. The color fill of cells corresponds to the land use colors in the Urban Atlas legend.}
\label{fig:cheltenham-fis}
\end{figure}

The most frequent displayed two-category combination is \{12100,11210\}---\emph{Industrial, commercial, public, military and private units} with \emph{Discontinuous dense urban fabric}---which occurs in 714 of 1,185 focal transactions (60.25\%). The combinations \{11210,11220\} and \{12100,11220\} occur in 55.95\% and 54.35\% of transactions, respectively, while the three-category combination \{12100,11210,11220\} occurs in 45.49\%. These values quantify recurrence across focal neighborhoods; they do not measure the area occupied by the corresponding classes or the number of same-class polygons within a neighborhood.

\begin{figure}[!htb]
\centering
\begin{minipage}[t]{0.4\linewidth}
\centering
\includegraphics[width=\linewidth]{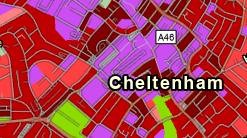}
\captionof{figure}{Central Cheltenham: mix of dense (11210) and medium-density urban fabric (11220) with commercial units (12100).}
\label{fig:central-cheltenham}
\end{minipage}\hfill
\begin{minipage}[t]{0.59\linewidth}
\centering
\includegraphics[width=\linewidth]{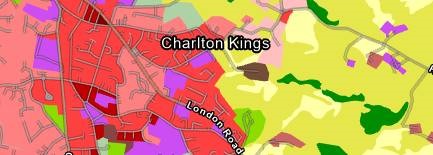}
\captionof{figure}{Charlton Kings (part of Cheltenham): discontinuous low-density (11230) and medium-density urban fabric (11220) coexist with commercial units (12100).}
\label{fig:charlton-kings}
\end{minipage}
\end{figure}

Figures~\ref{fig:central-cheltenham} and~\ref{fig:charlton-kings} illustrate how selected itemsets correspond to recognizable local configurations. The \{11220,11230\} combination of discontinuous medium- and low-density urban fabric has 27.93\% support, while \{23000,21000\}, combining pasture and arable land, has 24.39\% support. These examples also illustrate an interpretive limitation of the Urban Atlas nomenclature: class 12100 aggregates industrial, commercial, public, military, and private units, so a high-support itemset containing 12100 cannot by itself distinguish among those functions.

\subsection{Exact Feature Reconstruction and Input Validation}\label{sec:validation-results}

Across the 100 study areas, the analysis contains 290,396 focal-polygon transactions and 86,591 distinct within-city transaction patterns; 203,805 transaction occurrences are repetitions of patterns appearing elsewhere within the same city. The transaction rows contain 27 observed Urban Atlas codes and no repeated code within a final row, consistent with the set-based transaction representation described above.

At the exact 10\% support threshold, mining produces 17,847 nonzero city--itemset values and a union of 1,543 distinct itemsets. Independent exact re-mining reproduces the negFIN support counts for all 100 cities with zero discrepancies. The resulting 100-by-1,543 support matrix also agrees exactly with the matrix used for the cross-city analysis after alignment. These checks establish that the clustering and sensitivity analyses operate on a reproducible exact-support representation rather than on a reduced or rounded approximation.

\subsection{Original-Space Clustering and Choice of Seven Groups}\label{sec:groups}

\begin{figure}[!htb]
\centering
\includegraphics[width=\linewidth]{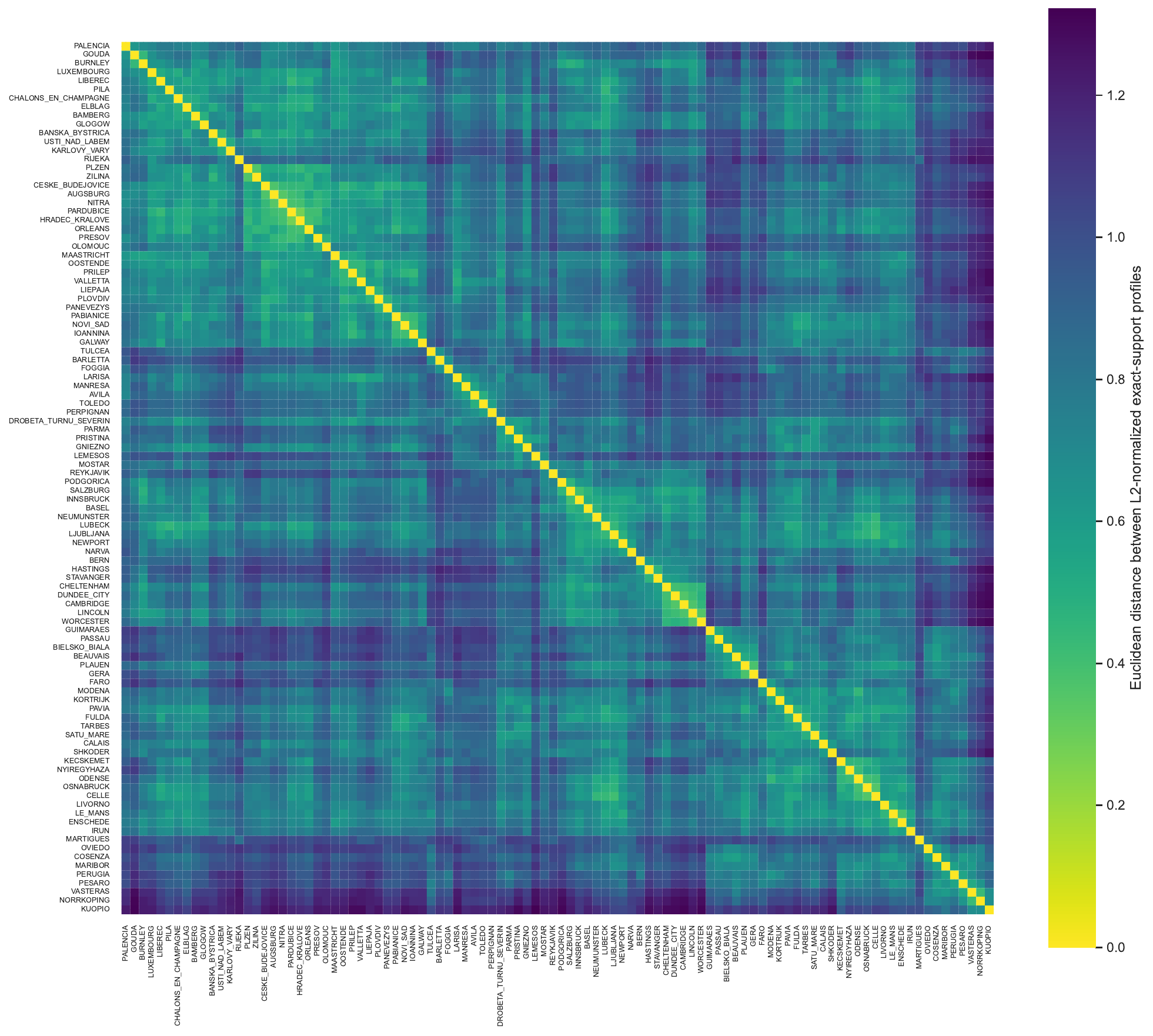}
\caption{Pairwise Euclidean distances between L2-normalized exact-support profiles in the original city-by-itemset feature space. Cities are ordered according to the primary Ward hierarchy.}
\label{fig:heatmap}
\end{figure}

Figure~\ref{fig:heatmap} shows pairwise Euclidean distances between L2-normalized exact-support profiles in the original 1,543-dimensional feature space. The block structure indicates substantial within-group similarity but also shows that the data form a graded rather than perfectly separated structure.

\begin{figure}[!htb]
\centering
\includegraphics[width=\linewidth]{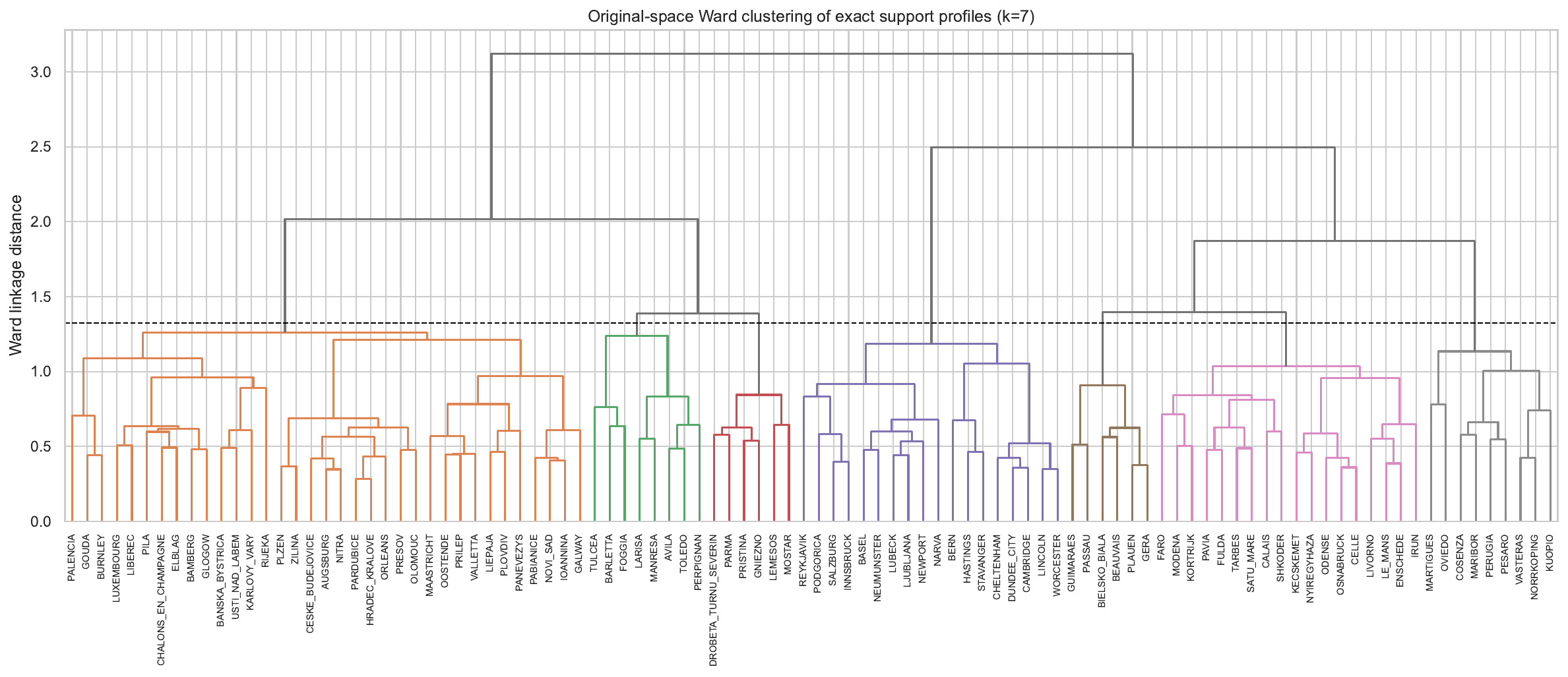}
\caption{Ward hierarchical clustering of the L2-normalized exact-support profiles in the original feature space. Colors show the retained seven-cluster descriptive partition.}
\label{fig:dendrogram}
\end{figure}

The Ward hierarchy is shown in Figure~\ref{fig:dendrogram}. Candidate values $k=2,\ldots,10$ do not produce a single unanimously preferred solution across all internal criteria. The highest Silhouette value occurs at $k=4$ (0.1265), whereas the highest Calinski--Harabasz value occurs at $k=2$ (15.9101). The seven-cluster solution has Silhouette 0.1213, Calinski--Harabasz 10.0405, and the lowest Davies--Bouldin value among the candidates (1.8355).

The paired 500-repetition feature-subsampling analysis further shows that $k=6$ and $k=7$ are effectively tied in stability: mean ARI is 0.761596 for $k=6$ and 0.760126 for $k=7$, a difference of only 0.00147. Under the declared equal-weight ranking of Silhouette, Calinski--Harabasz, Davies--Bouldin, and feature stability, $k=7$ obtains the best mean rank (3.50; $k=6$: 4.00). Seven clusters are therefore retained as a useful descriptive resolution that combines strong overall ranking, the best Davies--Bouldin value, near-leading feature stability, and planning-relevant differentiation. This choice is not interpreted as proof that seven is the uniquely correct number of city groups.

The final cluster sizes are 35, 8, 6, 18, 6, 18, and 9 cities. For interpretation, a medoid is defined as the city with the smallest total within-cluster Euclidean distance in the normalized feature space. Table~\ref{tab:cluster-profiles} summarizes the medoid and the itemset characteristics that most distinguish each cluster from the 100-city average.

\begin{table}[!htb]
\caption{Descriptive profiles of the seven original-space clusters. The non-artificial column is the median share of focal transactions containing at least one non-artificial land-use class; it is not a land-area share.}
\label{tab:cluster-profiles}
\centering
\scriptsize
\begin{tabularx}{\linewidth}{@{}c c l X c@{}}
\toprule
Cluster & $n$ & Medoid & Main distinguishing configuration & Non-artificial transaction share \\
\midrule
1 & 35 & České Budějovice & Continuous urban fabric with roads, dense fabric, and mixed artificial-surface uses & 0.659 \\
2 & 8 & Ávila & Herbaceous vegetation, arable land, and permanent-crop interfaces & 0.785 \\
3 & 6 & Priština & Land without current use with roads, dense fabric, and mixed urban uses & 0.680 \\
4 & 18 & Salzburg & Discontinuous medium-density fabric with roads and mixed artificial-surface uses & 0.617 \\
5 & 6 & Plauen & Discontinuous low-density fabric associated with pasture and forest & 0.831 \\
6 & 18 & Odense & Arable-land and pasture interfaces, often with roads and some industrial or isolated structures & 0.769 \\
7 & 9 & Västerås & Forest, isolated structures, and pasture; strongest non-artificial signal & 0.900 \\
\bottomrule
\end{tabularx}
\end{table}

Complete city-to-cluster memberships are provided in Supplementary Table~S2, with the corresponding city maps grouped by cluster in Supplementary Section~S3.

These labels are descriptive summaries of land-use co-presence profiles. They are not interpreted as causal planning-system categories, normative rankings, or universal city typologies.

\subsection{Sensitivity to Minimum Support and Analytical Representation}\label{sec:sensitivity-results}

Across 5--15\% minimum support, the union of retained itemsets decreases from 4,202 to 801, while pairwise city-distance rankings remain highly stable relative to the 10\% solution (Spearman 0.987--0.995; Table~\ref{tab:threshold-sensitivity}). Discrete membership is more sensitive at the endpoints (ARI 0.433 at 5\% and 0.591 at 15\%) than at 7.5\% and 12.5\% (0.810 and 0.911). We therefore treat 10\% as a pragmatic balance between retained detail and a tractable common representation, not as a universal optimum.

\begin{table}[!htb]
\caption{Sensitivity of the fixed seven-cluster solution to the exact minimum-support threshold.}
\label{tab:threshold-sensitivity}
\centering
\small
\begin{tabular}{@{}rrrr@{}}
\toprule
Minimum support (\%) & Union itemsets & ARI vs. 10\% & Distance Spearman vs. 10\% \\
\midrule
5.0  & 4202 & 0.433 & 0.989 \\
7.5  & 2370 & 0.810 & 0.995 \\
10.0 & 1543 & 1.000 & 1.000 \\
12.5 & 1087 & 0.911 & 0.995 \\
15.0 & 801  & 0.591 & 0.987 \\
\bottomrule
\end{tabular}
\end{table}

\begin{figure}[!htb]
\centering
\includegraphics[width=\linewidth]{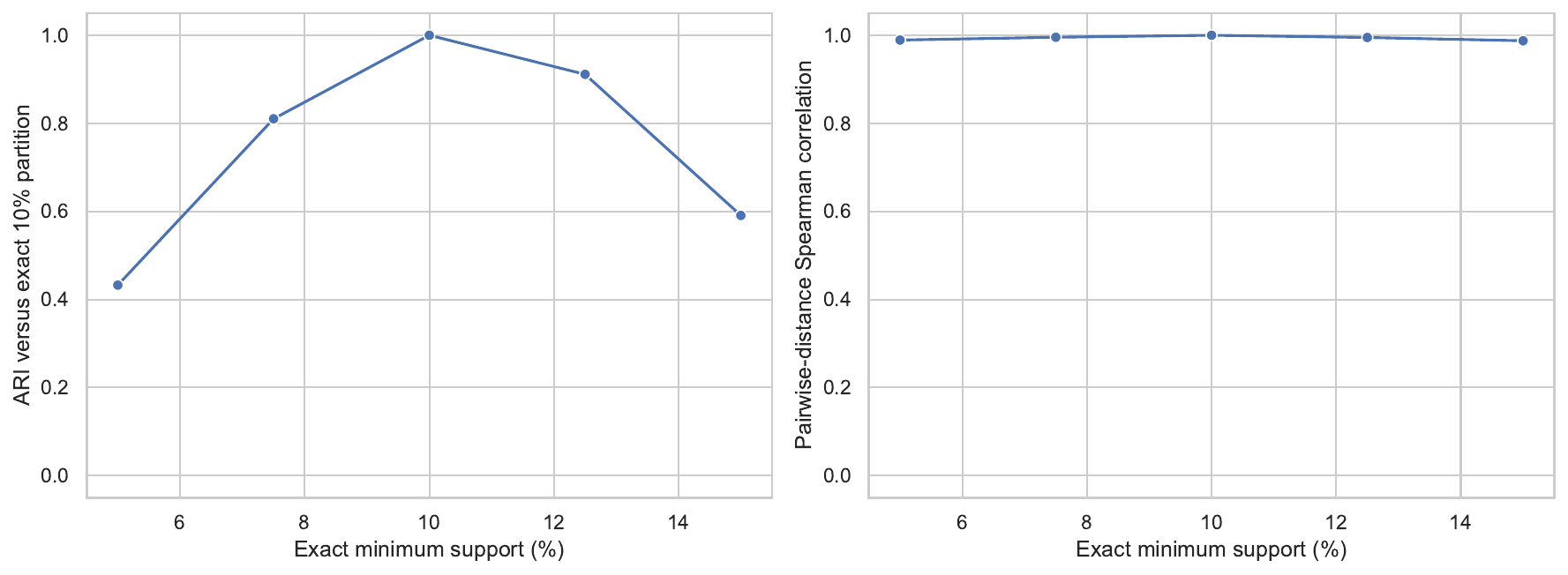}
\caption{Sensitivity of the seven-cluster partition and pairwise city-distance geometry to the exact minimum-support threshold. ARI measures agreement of discrete memberships with the 10\% solution; the Spearman coefficient compares pairwise distances.}
\label{fig:threshold-sensitivity}
\end{figure}

Representation choices affect discrete membership more strongly than broad distance geometry (Table~\ref{tab:representation-sensitivity}). Integer rounding has little effect (ARI 0.968), whereas binary itemset presence (0.286), artificial-only itemsets (0.291), and one vote per unique transaction pattern (0.302) substantially alter the partition. Removing row normalization gives ARI 0.860, PCA representations retaining 90\% and 95\% variance give 0.819 and 0.842, excluding road code 12220 gives 0.748, and cosine distance with average linkage gives 0.502. Repeated focal-neighborhood occurrences therefore contribute materially to the support profiles. Absolute support counts followed by rowwise L2 normalization reproduce the relative-support solution exactly because the two representations differ only by a city-specific scalar before normalization; this equivalence is not a test of within-neighborhood neighbor-count intensity.

\begin{table}[!htb]
\caption{Sensitivity to alternative representations and clustering geometries.}
\label{tab:representation-sensitivity}
\centering
\scriptsize
\begin{tabularx}{\linewidth}{@{}Xcc@{}}
\toprule
Alternative analysis & ARI vs. primary & Distance Spearman vs. primary \\
\midrule
Integer-rounded support, L2 + Ward & 0.968 & 1.000 \\
Binary itemset presence, L2 + Ward & 0.286 & 0.937 \\
Absolute support counts, L2 + Ward & 1.000 & 1.000 \\
Exact support without row normalization + Ward & 0.860 & 0.876 \\
Cosine distance + average linkage & 0.502 & 1.000 \\
PCA 90\% variance + Ward (31 components) & 0.819 & 0.999 \\
PCA 95\% variance + Ward (49 components) & 0.842 & 1.000 \\
Exclude itemsets containing road code 12220 & 0.748 & 0.997 \\
Artificial-only itemsets & 0.291 & 0.826 \\
One vote per unique transaction pattern & 0.302 & 0.860 \\
\bottomrule
\end{tabularx}
\end{table}

\subsection{UMAP as a Visualization and Diagnostic}\label{sec:umap-results}

\begin{figure}[!htb]
\centering
\includegraphics[width=\linewidth]{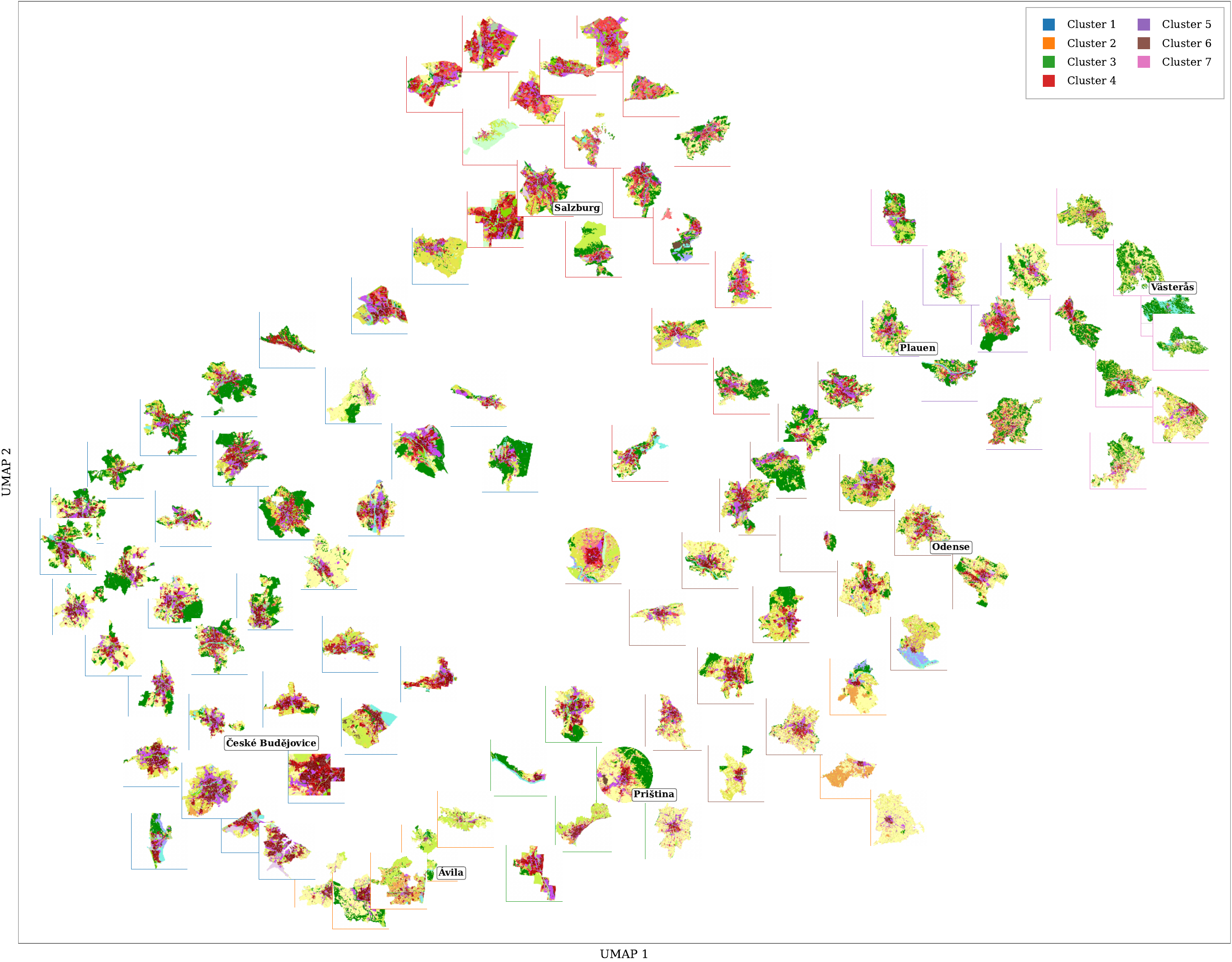}
\caption{Image-based UMAP visualization of the 100 cities. Each city is represented by its Urban Atlas land-use map, and the thumbnail border color indicates its fixed original-space Ward cluster. The labeled cities are cluster medoids. Small display-only positional adjustments reduce thumbnail overlap; these adjustments do not affect the underlying UMAP coordinates, cluster assignments, or diagnostic calculations. UMAP is used only for visualization and diagnostic assessment.}
\label{fig:umap-scatter}
\end{figure}

Figure~\ref{fig:umap-scatter} presents an image-based visualization of one UMAP embedding of the 100 cities. Each city is represented by its Urban Atlas land-use map, and the thumbnail border indicates membership in one of the seven fixed original-space Ward clusters. The seven cluster medoids are labeled to provide reference cases for interpreting the spatial arrangement of the groups. Small display-only adjustments are used where necessary to reduce thumbnail overlap; the underlying UMAP coordinates are retained unchanged for the diagnostic analyses. The resulting two-dimensional proximity is useful for visual exploration but is not the basis of the primary cluster assignments.

Across the 54 UMAP parameter/seed combinations, trustworthiness ranges from 0.854 to 0.914 and corrected nearest-10-neighbor overlap with the original space ranges from 0.516 to 0.597. Spearman correlation between original-space and two-dimensional pairwise distances ranges from 0.520 to 0.731. If Ward clustering is applied to the two-dimensional embeddings only as a diagnostic, agreement with the primary original-space labels is moderate rather than complete (ARI 0.342--0.612; NMI 0.598--0.758). These results support using UMAP as a visualization tool while retaining the full itemset-support feature space for the primary grouping.

\subsection{Influence of Non-Artificial Land-Use Content}\label{sec:rural-results}

Non-artificial land-use content is a substantive component of the cross-city comparison rather than a negligible background effect. Restricting the representation to artificial-surface itemsets produces ARI 0.291 relative to the primary solution and reduces pairwise distance-rank agreement to 0.826. This is one of the largest changes observed in the representation sensitivity analysis.

\begin{figure}[!htb]
\centering
\includegraphics[width=\linewidth]{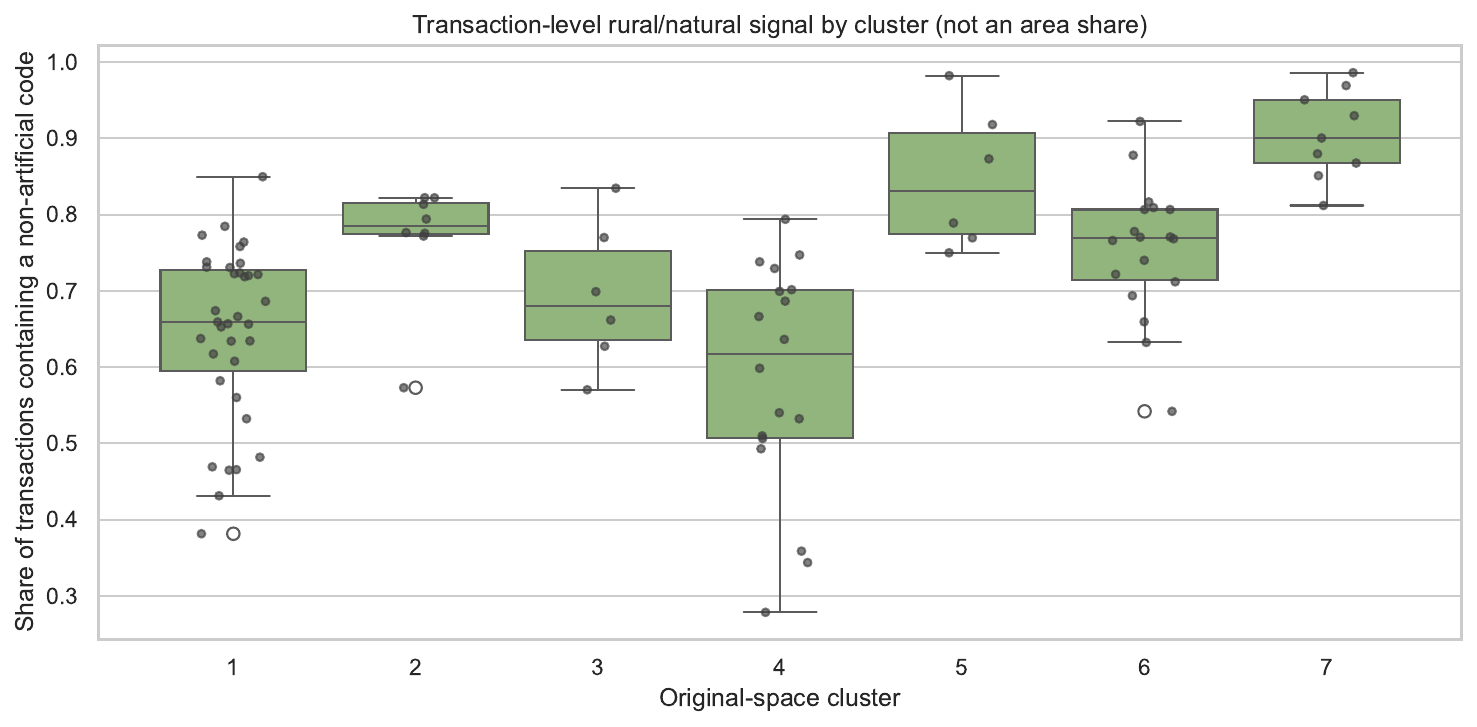}
\caption{Share of focal-polygon transactions containing at least one non-artificial Urban Atlas class, summarized by original-space cluster. The measure represents transaction prevalence, not land-area share.}
\label{fig:rural-signal}
\end{figure}

Figure~\ref{fig:rural-signal} shows the city-level share of focal transactions containing at least one non-artificial class. Median shares by cluster are 0.659, 0.785, 0.680, 0.617, 0.831, 0.769, and 0.900 for Clusters 1--7, respectively. These are transaction-prevalence measures, not proportions of land area.

Exploratory Kruskal--Wallis comparisons across the seven clusters yield $H=51.99$ for mean non-artificial classes per transaction (Holm-adjusted $p=5.62\times10^{-9}$; descriptive $\eta^2=0.566$), $H=50.03$ for the share of transactions containing a non-artificial class (adjusted $p=9.26\times10^{-9}$; $\eta^2=0.455$), and $H=15.07$ for the share of non-artificial-only transactions (adjusted $p=0.0197$; $\eta^2=0.224$). Because these metrics are constructed from the same transactions used to form the clusters, the tests are interpreted descriptively. Greater evidential weight is placed on the artificial-only re-clustering sensitivity.

Country and cluster membership are also associated descriptively (Cram\'er's $V=0.693$, 5,000-permutation $p=0.00020$). This pattern suggests geographic structuring that may motivate comparative planning investigation, but it does not demonstrate that national planning institutions or regulations caused the clusters because the model contains no institutional covariates and national samples are uneven.

\FloatBarrier

\section{Discussion}\label{sec:discussion}

The analysis shows that recurring local land-use co-presence provides a distinct basis for cross-city comparison. Rather than summarizing each city only through aggregate composition or morphology indicators, the representation records how often multi-category combinations recur across focal neighborhoods. The seven retained clusters therefore describe different prevalence profiles of local land-use interfaces. Cluster 1 is distinguished by continuous urban fabric and mixed artificial-surface co-presence, Cluster 4 by discontinuous medium-density fabric, Clusters 2 and 6 by agricultural and pastoral interfaces, Cluster 5 by low-density fabric with pasture and forest, Cluster 3 by land without current use in combination with urban uses, and Cluster 7 by forest, isolated structures, and pasture. These patterns complement established morphology- and indicator-based approaches to comparing European cities \citep{schwarz_urban_form_2010,prastacos_using_2017}.

For planning, the most defensible use of these groups is comparative rather than normative. A city can be placed alongside others that exhibit similar recurring land-use interfaces, providing candidate peer cases for more detailed investigation, benchmarking, or scenario comparison. This use is consistent with the role of comparative indicators in planning \citep{monteiro_benchmarking_cities_2024} but does not imply that cities in the same cluster share identical planning institutions, histories, or policy outcomes. The observed country--cluster association is therefore treated as a hypothesis generator. Institutional explanations would require explicit planning-system, regulatory, governance, and historical variables of the kind emphasized in comparative planning research \citep{schmidt_planning_institutions_2024}.

To illustrate how the descriptive clusters can inform planning interpretation without implying causation, selected cluster medoids were compared with their official local planning frameworks. These examples are intended as planning-oriented contextual illustrations rather than explanations of all seven clusters; the present analysis does not include harmonized institutional, regulatory, or policy variables for the 100 cities that would support systematic causal comparison of planning systems. Salzburg, the medoid of Cluster~4, is characterized in the present analysis by recurring discontinuous medium-density urban fabric together with roads and other artificial-surface uses. Salzburg's spatial-development framework emphasizes inward urban development, transformation of underused areas, and the coordination of housing and employment development with the protection of green and open spaces.\footnote{Stadt Salzburg, ``R{\"a}umliches Entwicklungskonzept (REK): R{\"a}umliche Entwicklungsziele,'' official municipal planning documentation, \url{https://www.stadt-salzburg.at/rek-neu/rek-textteil-raeumliche-entwicklungsziele}, accessed September~7, 2026.} This planning context provides a useful basis for interpreting the cluster's mixture of medium-density development and heterogeneous urban interfaces, although it does not establish that the planning framework caused the observed frequent-itemset profile.

Odense, the medoid of Cluster~6, illustrates a different type of interface. Its empirical profile is distinguished by recurring combinations of arable land, pasture, roads, and selected artificial uses. Odense's urban-development strategy includes a ``green ring'' intended to distinguish the contiguous city from surrounding settlements, protect landscape and nature, and structure future development primarily within the ring and designated development corridors.\footnote{Odense Kommune, ``Den gr{\o}nne ring,'' official municipal planning documentation, \url{https://www.odense.dk/gronring}, accessed September~7, 2026.} The prominence of agricultural and pastoral co-presence in Cluster~6 is therefore planning-relevant because it identifies locations where the relationship between built development and surrounding non-artificial land is an important component of the comparative urban profile.

V{\"a}ster{\aa}s, the medoid of Cluster~7, has the strongest non-artificial signal among the seven clusters and is characterized by recurring forest, pasture, and isolated-structure combinations. The municipality's comprehensive plan explicitly considers the long-term development of both the city and countryside, including the preservation and connection of natural and recreational areas through a green structure around the urban area.\footnote{V{\"a}ster{\aa}s stad, ``V{\"a}ster{\aa}s {\"o}versiktsplan,'' official municipal comprehensive-planning documentation, \url{https://www.vasteras.se/bygga-bo-och-miljo/kommunens-planarbete/oversiktsplan.html}, accessed September~7, 2026.} This example reinforces the interpretation of Cluster~7 as an urban--rural interface grouping rather than simply a low-urbanization residual category.

These examples demonstrate how the cluster profiles can be used to identify planning-relevant cases for subsequent contextual investigation. The correspondence between an empirical land-use profile and a planning document should not be interpreted as evidence that a particular planning policy produced the cluster assignment. Rather, the clustering provides a reproducible comparative starting point, while explanation of the processes that produced a city's spatial structure requires case-specific institutional, regulatory, historical, and socioeconomic evidence.

Sensitivity results reinforce a conditional interpretation of the seven groups. Pairwise distance geometry is highly stable across 5--15\% support, but endpoint thresholds and alternatives such as binary presence, unique-pattern weighting, cosine/average-linkage clustering, and removal of non-artificial itemsets can materially change membership. The clusters are therefore descriptive partitions conditional on analytical choices, while high distance-rank agreement under many alternatives indicates that the broader comparative structure is not driven by a single threshold or minor numerical choice. UMAP diagnostics further support retaining original-space clustering rather than defining groups in the two-dimensional embedding.

The unique-pattern sensitivity clarifies the duplicate-removal assumption. FIM treats each transaction as a set, so repeated occurrences of the same class within a neighborhood are collapsed and same-class multiplicity or geometric intensity cannot be recovered. Repeated configurations across different focal polygons, however, remain and contribute repeatedly to support; removing this recurrence changes the partition strongly (ARI 0.302). The representation therefore captures prevalence across neighborhoods but not within-neighborhood polygon counts, area, shared-boundary length, orientation, or exact distance.

Non-artificial content is also substantive: Clusters 5 and 7 have high transaction prevalence of non-artificial classes, and the artificial-only representation produces a markedly different partition. This supports treating agricultural, forest, pasture, and other non-artificial classes as meaningful urban--rural interface signals rather than background noise, consistent with peri-urban planning literature \citep{allen_periurban_interface_2003,hibbard_rurality_planning_2024}. Because Urban Core boundaries can include substantial non-artificial land and no paired Urban Core--full-FUA test is available, boundary definition remains a limitation.

The 100~m buffer is a study specification rather than a universally optimal planning distance. Historical Olomouc calibration shows that transaction breadth increases with buffer size, while 100~m retains a local neighborhood construct. Because equivalent multi-distance transactions are unavailable for all 100 cities, multi-city buffer robustness cannot be assessed here; future work should regenerate transactions from source polygons at several scales.

Urban Atlas nomenclature is another limitation: class 12100 combines industrial, commercial, public, military, and private units, so itemsets containing it cannot distinguish the contributing function. More detailed land-use and polygon-geometry data, together with demographic, socioeconomic, and qualitative planning evidence, would support stronger explanation of why similar co-presence profiles arise.

\section{Conclusion}\label{sec:conclusion}

This study develops a reproducible framework for comparing recurring local land-use co-presence across 100 selected European urban areas. The transaction database contains 290,396 focal neighborhoods, and exact frequent-itemset mining at 10\% support produces a common 1,543-feature representation. Ward clustering is performed directly on L2-normalized exact-support profiles in the original feature space, while UMAP is restricted to visualization and diagnostic assessment.

A seven-cluster descriptive solution is retained through a transparent multi-criterion evaluation that combines internal validity with 500 paired feature-subsampling repetitions. The results identify contrasts among continuous and discontinuous urban fabric, vacant or transitional land interfaces, agricultural and pastoral interfaces, and forest-dominant or isolated-structure contexts. Sensitivity analysis shows that pairwise city-similarity geometry remains highly stable across 5--15\% support thresholds, whereas exact cluster membership is more conditional on threshold endpoints and analytical representation. In particular, non-artificial land-use content and repeated focal-neighborhood occurrences materially influence the grouping.

For comparative planning, the framework supports candidate peer-city identification and investigation of recurring land-use interfaces rather than a fixed or normative European city typology. Interpretation remains conditional on the set-based transaction abstraction, the 100~m scale, Urban Core boundary definitions, Urban Atlas nomenclature, and the purposive sample. Future work should test multiple scales and boundary definitions, retain geometric and same-class intensity information where available, and integrate planning-system, demographic, socioeconomic, and qualitative evidence.

\section*{Data Availability Statement}

The datasets, processed inputs, analysis notebook, source code, and computational materials required to reproduce the results reported in this study are publicly available through the ULUMiner GitHub repository (\url{https://github.com/taiduydinh/ULUMiner}). The repository also documents the provenance of the data and the analytical workflow used in the study.

\section*{Acknowledgment} 

This paper was created within the project "Analysis, modelling, and visualization of spatial phenomena by geoinformation technologies V" (IGA\_PrF\_2026\_010) with the support of the Internal Grant Agency of Palacký University Olomouc).

\bibliographystyle{SageH}
\bibliography{_main}

\end{document}